\documentclass[english]{article}
\usepackage[T1]{fontenc}
\usepackage[latin1]{inputenc}
\usepackage{amsmath}
\usepackage{setspace}
\usepackage{amssymb}

\makeatletter
\newcommand{\lyxaddress}[1]{
\par {\raggedright #1
\vspace{1.4em}
\noindent\par}
}

\usepackage{babel}
\makeatother
\begin{document}

\title{Quaternion Analyticity of Time-Harmonic Dyon Field Equations }

\author{Jivan Singh$^{\text{(1)}}$, P. S. Bisht$^{\text{(2)}}$ and O. P.
S. Negi$^{\text{(2)}}$}

\maketitle
\begin{singlespace}

\lyxaddress{\begin{center}$^{\text{(1)}}$Department of Physics\\
 Govt. Post Graduate College \\
Pithoragarh- Uttarakhand, India\par\end{center}}

\lyxaddress{\begin{center}$^{\text{(2)}}$Department of Physics\\
Kumaun University\\
 S. S. J. Campus\\
Almora- 263601(Uttartakhand), India\par\end{center}}
\end{singlespace}

\lyxaddress{\begin{center}Email:-jgaria@indiatimes.com\\
 ps\_bisht123@rediffmail.com\\
 ops\_negi@yahoo.co.in\par\end{center}}

\begin{abstract}
Quaternion analysis of time dependent Maxwell's equations in presence
of electric and magnetic charges has been developed in unique, simple
and consistent manner. It has been shown that this theory is extended
consistently to time-harmonic Maxwell's equation for dyons.
\end{abstract}
~~~~~Dirac put forward the idea of magnetic monopole\cite{key-1}
while the fresh interests on the subject of monopoles was enhanced
by the work of t'' Hooft \cite{key-2} and Polyakov \cite{key-3}
and its extension to the case of dyons by Julia and Zee \cite{key-4}
. Consequently, these particles became intrinsic part of all current
grand unified theories \cite{key-5,key-6} with their enormous potential
importance \cite{key-7,key-8,key- 9,key-10,key-11,key-12,key-13}
in connection with various physical problems. Keeping in view the
results of Witten \cite{key-13} that monopoles are necessarily dyons,
a self-consistent and covariant quantum field theory for dyons each
carrying the generalized charges as its real and imaginary parts has
been constructed \cite{key-14,key-15} and accordingly the quaternionic
forms of generalized field equations of dyons are developed \cite{key-16,key-17,key-18,key-19}
in unique, simpler and compact notations. On the other hand Kravchenko
\cite{key-20} has analysed the Maxwell's equation in the presence
of sources, time-dependent electromagnetic fields in homogeneous (isotropic)
and chiral medium. In our previous papers, we \cite{key-21,key-22}
have analysed the generalized Maxwell's -Dirac equations in homogenous(isotropic)
medium, developed their quaternionic formulation and also obtained
the solutions for the classical problem of moving dyonsin simple,
compact and consistent manner. Keeping in view all these facts in
mind, in this paper, we have studied the Maxwell's equations in presence
of electric and magnetic sources (i.e. dyons) and extended the generaliged
electromagnetic field equations associated with dyons to the case
of time dependent harmonic Maxwell's-Dirac equations of dyons in simple,
compact and consistent way. It has been emphasized that the theory
reduces to the theory of dynamics of electric (magnetic) charge in
the absence of magnetic (electric) charge on dyons.

Assuming the existence of magnetic monopoles in homogeneous (isotropic)
medium, we have deduced the generalized Maxwell's-Dirac field equations
of dyons in the following form \cite{key-7};

\begin{eqnarray}
\overrightarrow{\nabla}.\overrightarrow{E} & = & \frac{\rho_{e}}{\epsilon}\nonumber \\
\overrightarrow{\nabla}.\overrightarrow{B} & = & \mu\rho_{m}\nonumber \\
\overrightarrow{\nabla}\times\overrightarrow{E} & = & -\frac{\partial\overrightarrow{B}}{\partial t}-\frac{\overrightarrow{j_{m}}}{\epsilon}\nonumber \\
\overrightarrow{\nabla}\times\overrightarrow{B} & = & \frac{1}{v^{2}}\frac{\partial\overrightarrow{E}}{\partial t}+\mu\overrightarrow{j_{e}}\label{eq:1}\end{eqnarray}
where $\rho_{e}$ and $\rho_{m}$ are respectively the electric and
magnetic charge densities while $\overrightarrow{j_{e}}$ and $\overrightarrow{j_{m}}$
are the corresponding current densities, $\overrightarrow{E}$ is
electric field intensity and $\overrightarrow{B}$ is magnetic field
intensity while $\epsilon$ and $\mu$ are defined respectively as
relative permitivity and permeability of the medium in electric and
magnetic fields. Differential equations (\ref{eq:1}) are the generalized
field equations of dyons in homogeneous (isotropic) medium and the
electric and magnetic fields are correspondingly called generalized
electromagnetic fields of dyons. These electric and magnetic fields
of dyons are expressed in terms of componets of two electromagnetic
potentials in following differential form in homogeneous (isotropic)
medium i.e.,

\begin{eqnarray}
\overrightarrow{E} & = & -\overrightarrow{\nabla}\phi_{e}-\frac{\partial\overrightarrow{C}}{\partial t}-\overrightarrow{\nabla}\times\overrightarrow{D}\label{eq:2}\\
\overrightarrow{B} & = & -\overrightarrow{\nabla}\phi_{m}-\frac{1}{v^{2}}\frac{\partial\overrightarrow{D}}{\partial t}+\overrightarrow{\nabla}\times\overrightarrow{C}\label{eq:3}\end{eqnarray}
where $\{ C^{\mu}\}=\{\phi_{e},\overrightarrow{vC}\}$ and $\{ D^{\mu}\}=\{ v\phi_{m},\overrightarrow{D}\}$
are the four-potentials respectively associated with electric and
magnetic charges of dyons. Let us define the complex vector field
$\overrightarrow{\psi}$in the following form \cite{key-7}

\begin{eqnarray}
\overrightarrow{\psi} & = & \overrightarrow{E}-iv\overrightarrow{B}.\label{eq:4}\end{eqnarray}
Equations (\ref{eq:2},\ref{eq:3}) and (\ref{eq:4}) then lead to
the following relation between generalized field and the components
of generalized four-potential as,

\begin{eqnarray}
\overrightarrow{\psi} & = & -\frac{\partial\overrightarrow{V}}{\partial t}-\overrightarrow{\nabla}\phi-iv(\overrightarrow{\nabla}\times\overrightarrow{V)}\label{eq:5}\end{eqnarray}
where $\{ V_{\mu}\}$ is the generalized four-potential of dyons in
homogeneous (isotropic) medium given by 

\begin{eqnarray}
V_{\mu} & = & \{\phi,\overrightarrow{V}\}\label{eq:6}\end{eqnarray}
where \begin{eqnarray}
\phi & = & \phi_{e}-iv\phi_{m}\label{eq:7}\end{eqnarray}
 and 

\begin{eqnarray}
\overrightarrow{V} & = & \overrightarrow{C}-i\,\frac{\overrightarrow{D}}{v}.\label{eq:8}\end{eqnarray}
Maxwell's field equation (\ref{eq:1}) in isotropic medium may then
be written in terms of generalized field $\overrightarrow{\psi}$as

\begin{eqnarray}
\overrightarrow{\nabla}\cdot\overrightarrow{\psi} & = & \frac{\rho}{\epsilon}\label{eq:9}\\
\overrightarrow{\nabla}\times\overrightarrow{\psi} & = & -iv(\mu\overrightarrow{J}+\frac{1}{v^{2}}\frac{\partial\overrightarrow{\psi}}{\partial t})\label{eq:10}\end{eqnarray}
where $\rho$and $\overrightarrow{J}$ are the generalized charge
and current source densities of dyons in homogeneous medium given
by \cite{key-7},

\begin{eqnarray}
\rho & = & \rho_{e}-i\frac{\rho_{m}}{v}\label{eq:11}\\
\overrightarrow{J} & = & \overrightarrow{j_{e}}-iv\overrightarrow{j_{m}}.\label{eq:12}\end{eqnarray}
In terms of complex potential the field equation is written as

\begin{eqnarray}
\square\phi & = & v\mu\rho\label{eq:13}\\
\square\overrightarrow{V} & = & \mu\overrightarrow{J}\label{eq:14}\end{eqnarray}
We may thus write the following tonsorial form of generalized Maxwell's
-Dirac equations of dyons in homogeneous (isotropic) medium \cite{key-7}
as

\begin{eqnarray}
F_{\mu\nu,\nu} & = & j_{\mu}^{e}\label{eq:17}\\
F_{\mu\nu,\nu}^{d} & = & j_{\mu}^{m}.\label{eq:18}\end{eqnarray}
Defining generalized field tensor of dyons as

\begin{eqnarray}
G_{\mu\nu} & = & F_{\mu\nu}-ivF_{\mu\nu}^{d}.\label{eq:19}\end{eqnarray}
One can directly obtain the following generalized field equation of
dyons in homogeneous (isotropic) medium i.e.

\begin{eqnarray}
G_{\mu\nu,\nu} & = & J_{\mu}.\label{eq:20}\end{eqnarray}
Using the Fourier transform, any electromagnetic field can be represented
as an infinite superposition of time-harmonic (monochromatic) fields.
These fields are normally cosidered as the main object of study in
radio electronics, wave propagation theory and many other branches
of physics and engineering. A time harmonic electromagnetic field
has the following form \cite{key-20},

\begin{eqnarray}
\overrightarrow{E}(x,t) & = & Re(\overrightarrow{E}(x)e^{-i\omega t})\label{eq:21}\end{eqnarray}
and

\begin{eqnarray}
\overrightarrow{B}(x,t) & = & Re(\overrightarrow{B}(x)e^{-i\omega t})\label{eq:22}\end{eqnarray}
where the electric field $\overrightarrow{E}$ and magnetic field
$\overrightarrow{B}$ depend on the spatial variables $x=(x_{1},x_{2},x_{3})$
and all dependence on time is contained in the factor $e^{-i\omega t}$.
$\overrightarrow{E}$ and $\overrightarrow{B}$ are expressed as the
complex vectors called the complex amplitudes of electromagnetic field
and $\omega$ denotes the frequency of oscillations.Substituting the
values of $\overrightarrow{E}$ and $\overrightarrow{B}$ into the
generalized dyonic equation (\ref{eq:1}) in isotropic medium, we
obtain

\begin{eqnarray}
\overrightarrow{\nabla}.\overrightarrow{E} & = & \frac{\rho_{e}}{\epsilon}\nonumber \\
\overrightarrow{\nabla}.\overrightarrow{B} & = & \mu\rho_{m}\nonumber \\
\overrightarrow{\nabla}\times\overrightarrow{E} & = & -i\omega\overrightarrow{B}-\frac{\overrightarrow{j_{m}}}{\epsilon}\nonumber \\
\overrightarrow{\nabla}\times\overrightarrow{B} & = & -\frac{i\omega}{v^{2}}\overrightarrow{E}+\mu\overrightarrow{j_{e}}.\label{eq:23}\end{eqnarray}
Let us denote $\alpha=\omega\sqrt{\epsilon\mu}=\frac{\omega}{v}$,
where the square root is chosen that $Im\alpha\geq0$, The quantity
$\alpha$ is called the wave number. Let us write the $\overrightarrow{D}$,
$\overrightarrow{E}$ and $\overrightarrow{B}$ in the following quaternionic
form as,

\begin{eqnarray}
D & = & \partial_{1}e_{1}+\partial_{2}e_{2}+\partial_{3}e_{3}\label{eq:24}\\
E & = & E_{1}e_{1}+E_{2}e_{2}+E_{3}e_{3}\label{eq:25}\\
B & = & B_{1}e_{1}+B_{2}e_{2}+B_{3}e_{3}\label{eq:26}\end{eqnarray}
where $e_{1},e_{2}$and $e_{3}$are the elements of aquaternion and
satisfy the following multiplication rule,

\begin{eqnarray}
e_{0}^{2} & = & 1\nonumber \\
e_{j}e_{k} & = & -\delta_{jk}+\epsilon_{jkl}e_{l}\label{eq:27}\end{eqnarray}
where $\delta_{jk}$ and $\epsilon_{jkl}$ (j, k, l= 1,2,3 and $e_{0}=1$)
are respectively the Kronecker delta and three-index Levi-Civita symbol.
Using equatios (\ref{eq:24},\ref{eq:25},\ref{eq:26}) , we get the
following quaternion differential equations i.e.

\begin{eqnarray}
D\overrightarrow{E} & = & (\partial_{1}e_{1}+\partial_{2}e_{2}+\partial_{3}e_{3})(E_{1}e_{1}+E_{2}e_{2}+E_{3}e_{3})\nonumber \\
= & -\frac{\rho_{e}}{\epsilon} & -\frac{\overrightarrow{j_{m}}}{\epsilon}+i\omega\overrightarrow{B}\label{eq:28}\end{eqnarray}
and

\begin{eqnarray*}
D\overrightarrow{B} & = & (\partial_{1}e_{1}+\partial_{2}e_{2}+\partial_{3}e_{3})(E_{1}e_{1}+E_{2}e_{2}+E_{3}e_{3})\end{eqnarray*}

\begin{eqnarray}
= & -\mu\rho_{m}+ & \mu\overrightarrow{j_{e}}-i\frac{\omega}{v^{2}}\overrightarrow{E}.\label{eq:29}\end{eqnarray}
Let us introduce the following pairs of purely vectorial biquaternionic
functions,

\begin{eqnarray}
\overrightarrow{l} & = & -\frac{i\omega}{v^{2}}\overrightarrow{E}+\alpha\overrightarrow{B}\label{eq:30}\end{eqnarray}
and 

\begin{eqnarray}
\overrightarrow{m} & = & \frac{i\omega}{v^{2}}\overrightarrow{E}+\alpha\overrightarrow{B}.\label{eq:31}\end{eqnarray}
Taking the divergence of third and fourth equation (\ref{eq:23}),
we get the following pairs of continuity equation for electric and
magnetic charges i.e.,

\begin{eqnarray}
\overrightarrow{\nabla}.\overrightarrow{j_{e}}-i\omega\rho_{e} & = & 0\label{eq:32}\end{eqnarray}

and 

\begin{eqnarray}
\overrightarrow{\nabla}.\overrightarrow{j_{m}}-i\omega\mu\epsilon\rho_{m} & = & 0.\label{eq:33}\end{eqnarray}
Applying the quaternionic operator $D$ given by equation (\ref{eq:24})to
the quaternionic form of $J$ the generalized current of dyons and
using equations (\ref{eq:28},\ref{eq:29} )and (\ref{eq:32},\ref{eq:33}),
we get

\begin{eqnarray}
D\overrightarrow{l} & = & \frac{1}{\epsilon v^{2}}[\overrightarrow{\nabla}.\overrightarrow{J}*]+\alpha\mu\overrightarrow{J}*+\alpha\overrightarrow{J}\label{eq:34}\end{eqnarray}
where $\overrightarrow{J}*$is the complex conjugate of dyonic current
density in homogeneous (isotropic) medium given by equation (\ref{eq:12}).
Thus $\overrightarrow{J}$ satisfies the equation which is derived
by equation (\ref{eq:34}) as,

\begin{eqnarray}
(D-\alpha)\overrightarrow{J} & = & \mu[\overrightarrow{\nabla}.\overrightarrow{J}*]+\alpha\mu\overrightarrow{J}*.\label{eq:35}\end{eqnarray}
Analogous to equation (\ref{eq:35}), $\overrightarrow{m}$ also satisfies
the equation (\ref{eq:35}) as,

\begin{eqnarray}
(D-+\alpha)\overrightarrow{m} & = & -\mu[\overrightarrow{\nabla}.\overrightarrow{J}]+\alpha\mu\overrightarrow{J}.\label{eq:36}\end{eqnarray}
Thus, the process of diagonalization can be written in the matrix
form as 

\begin{eqnarray}
\left(\begin{array}{cc}
D & -i\omega\\
i\omega & D\end{array}\right)\left(\begin{array}{c}
\overrightarrow{E}\\
\overrightarrow{B}\end{array}\right) & = & B_{\alpha}\left(\begin{array}{cc}
D-\alpha & 0\\
0 & D+\alpha\end{array}\right)B_{\alpha}^{-1}\left(\begin{array}{c}
\overrightarrow{E}\\
\overrightarrow{B}\end{array}\right)\label{eq:37}\end{eqnarray}
where

\begin{eqnarray}
B_{\alpha} & = & \left(\begin{array}{cc}
-\frac{i\omega}{v^{2}} & \alpha\\
\frac{i\omega}{v^{2}} & \alpha\end{array}\right)\label{eq:38}\end{eqnarray}

and

\begin{eqnarray}
B_{\alpha}^{-1} & = & \left(\begin{array}{cc}
-\frac{v^{2}}{i\omega} & \frac{v^{2}}{i\omega}\\
\frac{1}{\alpha} & \frac{1}{\alpha}\end{array}\right).\label{eq:39}\end{eqnarray}
As such, we have obtained the two decoupled equations for the unknown
vectors $\overrightarrow{J}$ and $\overrightarrow{m}$ , which simplifies
the analysis of the generalized Maxwell's-Dirac equation of dyons
in homogeneous (isotropic) medium. These equations reduce to the theory
of electric charge (magnetic) monopole predicted earlier by Kravchenko
\cite{key-20} in the absence of magnetic (electric charge) or vice
versa.

\end{document}